\pdfoutput=1

\documentclass[11pt]{article}

\usepackage[]{emnlp2021}

\usepackage{times}
\usepackage{latexsym}
\usepackage{amsfonts}
\usepackage{amsmath}
\usepackage{booktabs}
\usepackage{multirow}
\usepackage{graphicx}
\usepackage{enumitem}

\usepackage[hang,flushmargin]{footmisc}

\usepackage[T1]{fontenc}

\usepackage[utf8]{inputenc}

\usepackage{microtype}
\usepackage{inconsolata}

\title{Evaluating Token-Level and Passage-Level Dense Retrieval Models for Math Information Retrieval}

\author{Wei Zhong, Jheng-Hong Yang, Yuqing Xie, \and Jimmy Lin \\[1ex]
  David R. Cheriton School of Computer Science\\
  University of Waterloo \\[1ex]
  \texttt{\{w32zhong,jheng-hong.yang,yuqing.xie,jimmylin\}@uwaterloo.ca}
}

\begin{document}
\maketitle
\begin{abstract}
With the recent success of dense retrieval methods based on bi-encoders, studies have applied this approach to various interesting downstream retrieval tasks with good efficiency and in-domain effectiveness.
Recently, we have also seen the presence of dense retrieval models in Math Information Retrieval (MIR) tasks,
but the most effective systems remain classic retrieval methods that consider hand-crafted structure features.
In this work, we try to combine the best of both worlds:\ a well-defined structure search method for effective formula search and efficient bi-encoder dense retrieval models to capture contextual similarities.
Specifically, we have evaluated two representative bi-encoder models for token-level and passage-level dense retrieval on recent MIR tasks.
Our results show that bi-encoder models are highly complementary to existing structure search methods, and we are able to advance the state-of-the-art on MIR datasets.
\end{abstract}

\section{Introduction}

Math Information Retrieval (MIR) is a special information retrieval domain that deals with heterogeneous data.
The core task in this field is to retrieve relevant information from documents that contain math formulas.
As digitized math content (mostly in \LaTeX{} markup or MathML format) becomes readily available nowadays,
being able to index and retrieve formulas (or equations) in those documents effectively is possibly one of the hard nuts left to be cracked before we can search freely for scientific documents, educational materials, and other math content.

The need to measure similarities for highly structured formulas with special semantic properties and to model their connections to the surrounding text have a few interesting consequences:
(1) Heuristic scores like the term-frequency factor in tf--idf scoring variants become less relevant in formula similarity assessment because symbols in a math formula can be interchangeable and similarity may depend on expression structure rather than frequency of co-occurrence. 
(2) At the same time, the same math content can be expressed differently, e.g., $\{1,2,...\}$ and $\mathbb{N}^{+}$ represent the same concept but they are made of totally different tokens. 
We also need to capture similar math expressions with different structure due to math transformations, e.g., $1+\frac 1x$ and $\frac{1 + x}{x}$. 
These have made structure search approaches alone suboptimal.
(3) Many existing methods score math and text separately because they are of different modalities; however, failing to catch cross references between text and math will penalize retrieval effectiveness.
For example, a top effective math-aware search engine adopting traditional ad-hoc search techniques~\cite{fraser2018tangentL,ng2020dowsing, ng2021dowsing} tunes a hyperparameter to weight text and formulas in two separate passes, which provides little awareness of the connections between formulas and their surrounding text.
The aforementioned challenges have put limitations on further advances in this field.

On the other hand, recent bi-encoder dense retrieval models~\cite{karpukhin2020dpr,santhanam2021colbertv2,hofstatter2021TAS_B,formal2021spladeV2,gao2021condenser} have been shown to be highly effective for in-domain retrieval while remaining efficient for large corpora in practice.
Compared to traditional retrieval methods, these models use dual deep encoders, usually built on top of a Transformer encoder architecture~\cite{vaswani2017attention,devlin2019bert}, to encode query and document passages separately and eventually output contextual embeddings.
Similarity scores can be efficiently computed given these embeddings, which limits costly neural inference to indexing time.
The effectiveness of these models can be attributed to the encoder's ability to capture contextual connections or even high-level semantics without the necessity for exact lexical matching.
This very complementary benefit compared to more rigorous structure search methods motivates us to investigate whether dense retrieval models can improve MIR results when combined with existing structure search methods.
We summarize the contributions of this work as follows:

\begin{itemize}[leftmargin=*]
  \item We have performed a fair effectiveness comparison of a token-level and a passage-level dense retrieval baseline in the MIR domain. To our knowledge, this is the first time that a DPR model has been evaluated in this domain.
  
  \item We have successfully combined dense retrievers with a structure search system and have been able to achieve new state-of-the-art effectiveness in recent MIR datasets.

  \item A comprehensive list of dense retrievers and strong baselines for major MIR datasets are covered and compared. We believe our well-trained models and data pipeline\footnote{Our model checkpoints and source code are made publicly available: \url{https://github.com/approach0/math-dense-retrievers/tree/emnlp2022}} can serve as a stepping stone for future research in this domain, which suffers from a scarcity of resources.
  
\end{itemize}

\section{Background and Related Work}

\subsection{Classic and Structure Search}

Research on math information retrieval started with the DLMF project from NIST decades ago~\cite{miller2003technical}.
Naturally, early studies~\cite{miller2003technical,youssef2005issue_and_methods} directly converted math symbols to textualized tokens (e.g., ``$+$'' will be converted to ``plus'') so they can be easily retrieved with existing IR systems.
Later, a line of studies~\cite{hijikata2009search,sojka2011art,lin2014mathematics,zanibbi2015tangent,kristianto2016mcat,fraser2018tangentL} utilizing full-text search engines additionally introduced various intermediate tree representations to extract features that capture more structure information.

The MathDowsers system~\cite{fraser2018tangentL,ng2020dowsing,ng2021dowsing,ng2021thesis} stands out in retrieval effectiveness by the incorporation of a mature full-text search engine and a curated list of over 5 types of features extracted from the Symbol Layout Tree (SLT) representation~\cite{zanibbi2012survey}.
Other features like the leaf-root paths extracted from the Operator Trees or representational MathML DOMs are also popular among researchers~\cite{hijikata2009search,yokoi2009approach,zhong2015thesis,zhong2016opmes}; these features are invariant to operand position mutation (e.g., due to commutativity) and require less storage.
More strict and top-down approaches~\cite{kohlhase2008mathwebsearch,schellenberg2012layout,zanibbi2016multi,zhong2019structural,mansouri2020tangentcfted} have also been proposed by evaluating well-defined math formula structure similarity or edit distance, resulting in higher precision in top-ranked results generally.
Furthermore, \citet{zhong2020accelerating} have shown that a top-down structure search method can be accelerated to achieve practically efficient first-stage retrieval as well.

\subsection{Data-Driven Methods}

More recently, data-driven approaches that incorporate word embeddings~\cite{gao2017preliminary,mansouri2019tangentcft}, GNNs~\cite{song2021GNN}, or Transformer models~\cite{peng2021mathbert,reusch2021albert,reusch2021tu_dbs} have also been proposed for the MIR domain.
By observing token co-occurrence and structure features during training, these models can discover synonyms or high-level semantic similarities, making them a good enhancement to strict structure matching.
However, previous Transformer-based retrievers in this domain~\cite{mansouri2021dprl,reusch2021albert,reusch2021tu_dbs} either only evaluate partial collections due to the adoption of expensive cross encoders, or cover only a token-level bi-encoder retriever, i.e., using the ColBERT model~\cite{khattab2020colbert,santhanam2021colbertv2}.
The effectiveness of a fine-tuned bi-encoder Transformer retriever for passage-level semantic similarity remains unknown.

In this work, we examine the DPR model \cite{karpukhin2020dpr} as a passage-level dense retriever baseline for the MIR domain.
We also fine-tune a ColBERT model~\cite{khattab2020colbert} that greatly outperforms the same type of models previously described in this domain.
Furthermore, previous efforts~\cite{mansouri2019tangentcft,peng2021mathbert} to consider structure features using data-driven models have achieved good levels of effectiveness; we will follow this path and evaluate the combination of structure-matching methods and dense retrieval.

Finally, some previous effective cross-encoder math retrieval runs~\cite{reusch2021tu_dbs} are based on further-pretrained backbone models in this domain.
However, this \textit{domain-adaptive pretraining} (DAPT)~\cite{gururangan2020dontstoppretrain} shows inconsistent benefits to downstream tasks~\cite{zhu2021notalwayshelp}.
In this work, we wish to investigate and compare different bi-encoder backbones on downstream retrieval effectiveness in a fair manner.

\subsection{Dense Retrieval Models}
\label{sect:models}
\smallskip \noindent
\textbf{DPR} \ 
In the Dense Passage Retriever (DPR) architecture~\cite{karpukhin2020dpr}, a Transformer encoder $E(\cdot)$ is applied to the query or passage:\ the output embedding corresponding to the \texttt{[CLS]} token is used to calculate a similarity score.
To facilitate retrieval efficiency, a simple dot product is used:
\begin{align}
S(q, p) = E(q)^T \cdot E(p)
\end{align}
\noindent where $S(q, p)$ represents the similarity between a query $q$ and a passage $p$.

During training, a pretrained model is used as the initial encoder state, and the encoder is optimized through the objective of a contrastive loss consisting of a query and a pair of positive and negative passages, $p^+$ and $p^-$.
A common practice in training a batch of queries $\{q_i\}^B_1$ is to utilize passages of other training instances from the batch as additional  \textit{in-batch negatives} in the loss function:
\begin{align}
\begin{split}
& \mathcal{L}^{(i)}(q_i, p^+_i, p^-_{i,1}, ..., p^-_{i,2B - 1}) \\
& = - \log \dfrac{\exp\left(S(q_i, p^+_i)\right)}{\exp\left(S(q_i, p^+_i)\right) + \displaystyle\sum_{j=1}^{2B - 1} \exp\left(S(q_i, p^-_{i,j})\right)}
\end{split}
\end{align}

\smallskip \noindent
\textbf{ColBERT} \ Instead of using a single passage-level embedding, the ColBERT model~\cite{khattab2020colbert,santhanam2021colbertv2} preserves all output embeddings for the similarity calculation.
Since each Transformer encoder is pretrained using the MLM objective~\cite{devlin2019bert}, the model provides fine-grained contextualized semantics for individual tokens.

Given a query token sequence $q = q_0, q_1, ... q_l$ and a passage token sequence $p = d_1, d_2, ... d_n$, ColBERT uses either dot product or L2 distance of normalized embeddings for computing the token-level similarity score $s(q_i, d_j)$.
During scoring, it locates the highest-scoring token of a passage $d_j$ for each query token $q_i$ (i.e., the MaxSim operator), and a summation is taken over these partial scores as the overall similarity between query $q$ and passage $p$:
\begin{align}
\label{eq:colbert}
S(q, p) = \sum_{i \in [E(q)]} \max_{j \in [E(p)]} s(q_i, d_j).
\end{align}
Similar to the DPR model, given a triple of query and a contrastive passage pair, i.e., $(q, p^+, p^-)$, the ColBERT model optimizes a pairwise softmax cross-entropy loss.

Because ColBERT uses all passage token embeddings, it applies a linear pooling layer on top of its backbone encoder to obtain smaller fixed-size ($d=128$ by default) embedding outputs for more efficient score computation or token-level indexing.
In addition, the model prepends two different special tokens, \texttt{[Q]} or \texttt{[D]}, to distinguish the encoding of a query or a passage.
In practice, the authors also demonstrate improved effectiveness via \textit{query augmentation} by rewriting \texttt{[PAD]} query tokens to \texttt{[MASK]} tokens before query encoding.

In end-to-end retrieval, however, ColBERT typically relies on two-stage query processing for efficiency:\
(1) A candidate set of tokens is retrieved using highly efficient approximate nearest neighbors (ANN) search techniques (e.g., \citealp{jegou2011IVFADC_R}).
(2) Then, the passages containing these tokens need to be located.
(3) Finally, the candidate passages are then sent to the GPU for fast matrix multiplication to calculate token similarities for each query and passage in the candidate pool.
Due to the candidate selection pipeline, this process is regarded as an approximate version of retrieval for the similarity search specified by  Eq.~\ref{eq:colbert}.

\subsection{Fusing Dense and Structure Signals}

Although \citet{peng2021mathbert} have performed structure mask pretraining for better matching formula substructures, their method is still based on additional structure embeddings generated from a different system.
However, we argue that a dense retrieval model may excel at adding fuzziness and recall to math retrieval without being constrained to require a structure match in candidates.
Given that previous math retrieval systems~\cite{zhong2020accelerating,zhong2021aa} have already incorporated effective formula structure matching, we wish to combine existing well-defined structure similarity search systems with more fuzzy and higher-level semantic search capabilities from dense retrieval models.

\section{Evaluation Setup}

\subsection{Datasets}

Evaluations in this paper are conducted on two recent MIR tasks:

\smallskip \noindent
\textbf{NTCIR-12 Wiki-Formula}~\cite{zanibbi2016ntcir} \ A formula-only retrieval task made from math-related pages in Wikipedia.
Both queries and documents are isolated formulas encoded using \LaTeX{}.
We consider all 20 concrete queries (no wildcards for formula variables) and index all (around 591,000) formulas as documents in this task.
Judgment ratings are provided on a scale of 0 to 3.
For each judged formula,
the ratings are mapped to
\textit{fully relevant} ($\ge 2$), \textit{partially relevant ($\ge 1$)}, or \textit{irrelevant} ($= 0$).

\smallskip \noindent
\textbf{ARQMath-2 (main task)}~\cite{mansouri2021arqmath2review} \ A CLEF answer retrieval task for math-related questions.
The collection includes roughly 1 million questions, containing 28 million formulas extracted from the MSE (Math StackExchange) website.\footnote{\url{https://math.stackexchange.com}}
There are 100 question posts sampled from MSE, where 71 of these questions are sufficiently evaluated (an average of 450 answers per topic are assessed by human experts). 
The official evaluation measurements in ARQMath are prime-versions of NDCG, MAP, and Precision at 10.
They differ from the original metrics in that unjudged documents from the ranked lists are removed before evaluation.
Relevance levels include \textit{High} ($=3$), \textit{Medium} ($=2$), \textit{Low} ($=1$), and \textit{Irrelevant} ($=0$).
High and Medium relevance are collapsed for binary evaluation metrics.

\medskip \noindent We use the official evaluation metrics and protocols for both tasks.
Each run contains a ranked list of 1000 documents per query.

\subsection{Pretraining Configurations}

We consider three types of pretrained Transformer backbones for downstream math retrieval tasks.

\smallskip \noindent
\textbf{BERT}~\cite{devlin2019bert} \  A Transformer encoder pretrained using MLM and NSP objectives on a large corpus comprising the Toronto Book Corpus and English Wikipedia.

\smallskip \noindent
\textbf{SciBERT}~\cite{beltagy2019scibert} \  A further pretrained Transformer encoder built on the BERT base model using 1.14M scientific papers with additional vocabularies for scientific content.

\begin{figure}
\small
\centering
\renewcommand{\arraystretch}{1.5}
\begin{tabular}{p{2.9in}}
\toprule
\textbf{Example: } Inequality between norm 1, norm 2 and norm $\infty$ of matrices:
$\|A\|_2 \leq \sqrt{ \|A\|_1 \|A\|_\infty}$
\\ \midrule
\textbf{Output: } Inequality between norm 1, norm 2 and norm <infty> of matrices: <Vert> <A> <Vert> <subscript> <2> <le> <root> <\{> <Vert> <A> <Vert> <subscript> <1> <Vert> <A> <Vert> <subscript> <infty> <\}>
\\ \bottomrule
\end{tabular}
\caption{\label{fig:pretokenize}
Example of pre-tokenized math content containing \LaTeX{} markup.
The output has all meaningful or syntactic \LaTeX{} tokens preserved.}
\end{figure}

\smallskip \noindent
\textbf{Our further-pretrained BERTs} \ 
We further pretrain the BERT base model on 1.69M math-related documents composed of texts and math formulas using the MLM and NSP objectives proposed by~\citet{devlin2019bert}.
Specifically, we crawl MSE and the Art of Problem Solving\footnote{\url{https://artofproblemsolving.com/community}} websites.
Out of 9M sentences from these documents, we extracted 2.2M sentence pairs for training.
All \LaTeX{} markup are pre-tokenized using the PyA0 toolkit~\cite{zhong2021pya0}, which unifies semantically identical tokens (e.g., \texttt{\textbackslash frac} and \texttt{\textbackslash dfrac}), adding 539 new tokens into the vocabulary space.
An example of the pre-tokenizing process can be found in Figure~\ref{fig:pretokenize}.
We treat these \LaTeX{} tokens as regular text during training.
Our own backbone is trained on eight A100 GPUs with a batch size of 240 for 3 and 7 epochs.

\subsection{Fine-Tuning Configurations}

On top of different backbones, we fine-tune our bi-encoder models using the ARQMath collection as training data (we use Q\&A posts prior to the year 2018).
Given a query $q$, we sample a positive passage $p^+$ from accepted answers, duplicate questions, or any answer posts to the query receiving more than 7 upvotes.
A random answer passage related to the same tags is treated as a hard negative sample $p^-$.
We obtain 607K $(q,p^+,p^-)$ triplets for training dense retrieval models.

We use the AdamW optimizer~\cite{loshchilov2017adamW} in all our experiments, with a weight decay of $0.01$, and a learning rate of $1 \times 10^{-6}$ for ColBERT and $3 \times 10^{-6}$ for DPR.
Following \citet{reusch2021tu_dbs}, we set the maximum number of input tokens to 512.

\smallskip \noindent
\textbf{DPR models trained for 1 epoch} \  To validate the effectiveness of our pretrained backbones, we design a comparative experiment where DPR models on different backbones are trained for the same number of steps ($\sim$550K iterations, approximately one epoch) with a batch size of 15.
The goal of these conditions is to quickly compare the effectiveness of different backbones.

\smallskip \noindent
\textbf{Fully-trained models} \ To maximize effectiveness, we fine-tune our DPR and ColBERT models on our backbone that has been further pretrained for 7 epochs.
We fine-tune DPR for 10 epochs with a batch size of 36, and ColBERT with a batch size of 30 for 3 epochs, both on A6000 GPUs.

\subsection{Structure Search Fusion}

The best way to add structure similarity awareness to dense retrieval models remains an important and open problem.
In this work, we make the first attempt to simply merge dense retrieval results with results generated from a structure search system, Approach0~\cite{zhong2019structural,zhong2020accelerating}.
Approach0 takes a top-down approach to evaluate formula similarities, and thus it is very complementary to more fuzzy semantic retrieval.

We evaluate search fusion against the NTCIR-12 and ARQMath-2 tasks.
Based on structure search results, which are generated by Approach0 and tuned on a different dataset (i.e., ARQMath-1), we perform one of two alternatives:\ (1) rerank the baseline using inference scores from DPR or ColBERT;
(2) linearly combine scores from the baseline and from DPR or ColBERT.
In the second case, we perform 5-fold cross validation to tune a weight $\alpha \in \{0.1, ..., 0.9\}$.
The final fusion score $S_f$ is interpolated by a convex combination:
$$
S_f = \alpha \cdot S_d + (1 - \alpha) \cdot S_a \text{,}
$$
\noindent where $S_d, S_a$ are the scores from the dense retrievers and the structure search, respectively.
Original scores are rescaled using min-max normalization, and when a document is missing from the other source during fusion, we set its score to zero.

\begin{table*}
\centering
\caption{\label{tab:dense}
Effectiveness comparisons of bi-encoder Transformers.
Rows (1)--(4) show DPR models fine-tuned for one epoch, starting from different pretrained backbones.
Our fully-trained passage-level and token-level models (DPR and ColBERT, respectively), rows (11) and (12), are compared with existing Transformer models in rows (5)--(10).
*/** denotes that the compared row performs weaker than the bottom row in each block, i.e., the 1-epoch fine-tuned and 7-epoch further-pretrained BERT in row (4) or our fully-pretrained ColBERT model in row (12), at $p<$ 0.05/0.01 level using the two-tailed pairwise $t$-test.
Underlined scores are not involved in any test of significance due to unavailable run files.
}
\resizebox{\textwidth}{!}{
\begin{tabular}{l|lll|lllll}
\toprule
    \multirow{2}{*}{\bf Runs} &
    \multicolumn{3}{c|}{\bf NTCIR-12 Wiki-Formula} &
    \multicolumn{5}{c}{\bf CLEF ARQMath-2} \\
    &
    Full BPref & Part. BPref & Judged ‰ &
    NDCG' & MAP' & P'@10 & BPref & Judged ‰ \\
\midrule
\bf DPR models fine-tuned for 1 epoch &&&& \\
(1) BERT~\citeyearpar{devlin2019bert} &0.505&0.393&21.9&0.174*&0.051&0.116&0.073&45.3 \\
(2) SciBERT~\citeyearpar{beltagy2019scibert} &0.512&0.363&21.1&0.176*&0.056&\bf 0.134&0.073&42.3 \\
(3) further-pretrained BERT (3 epochs)& 0.486&0.392&21.8&0.195&0.058&0.126&0.073&45.0 \\
(4) further-pretrained BERT (7 epochs) &\bf 0.522&\bf 0.439&23.2&\bf 0.200&\bf 0.060&0.130&\bf 0.081&46.8 \\[0.75ex]
\midrule
\bf Other bi-encoder Transformers &&&& \\
(5) MathBERT~\dag\ \citeyearpar{peng2021mathbert} &\bf \underline{0.614} &\bf \underline{0.736}&-&-&-&-&-&- \\
(6) TU\_DBS ColSciBERT~\citeyearpar{reusch2021tu_dbs} &-&-&-&0.028**&0.004**&0.009**&0.009**&17.5 \\
(7) TU\_DBS ColBERT~\citeyearpar{reusch2021tu_dbs}
&-&-&-&\underline{0.183}&\underline{0.053}&\underline{0.110}&-&- \\
(8) TU\_DBS ColARQBERT~\citeyearpar{reusch2021tu_dbs} &-&-&-&\underline{0.225}&\underline{0.073}&\underline{0.131}&-&- \\
(9) MIRMU CompuBERT~\citeyearpar{novotny2021ensembling} &-&-&-&0.262**&0.083**&0.135**&0.087**&69.2 \\
(10) FormulaEmb~\citeyearpar{indian2021bert} &-&-&-&\underline{0.161}&\underline{0.059}&\underline{0.197}&-&- \\[0.75ex]
\bf Our fully-trained models &&&& \\
(11) Our DPR &0.516 & 0.427 &23.4 & 0.270**&0.087**&0.152**&0.097**&66.3 \\
(12) Our ColBERT &0.545&0.483 & 25.1&
\bf 0.329	&\bf 0.128	& \bf 0.213	& \bf 0.136	& 69.5 \\
\bottomrule
\multicolumn{4}{l}{\dag: Ensemble system.}
\end{tabular}
}
\end{table*}

\begin{figure*}
  \centering
  \hspace*{-0.5in} \includegraphics[width=1\linewidth]{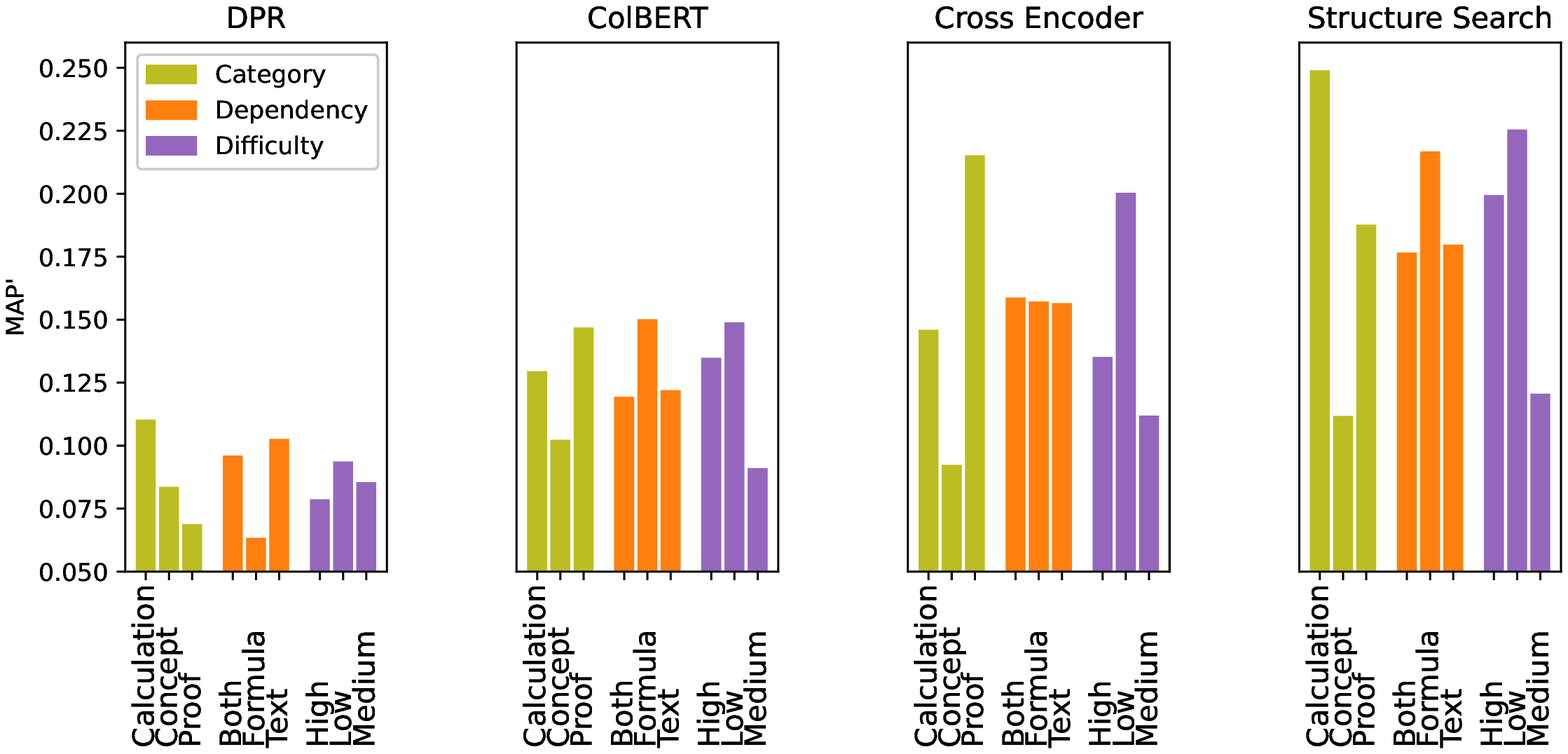}
  \vspace{-0.1in}
  \caption{The MAP$'$ scores produced by our fully-trained DPR, ColBERT, a cross encoder represented by the TU\_DBS primary run, and the structure matching retriever Approach0, all evaluated on the ARQMath-2 dataset.
  Results are divided by different topic categories (Calculation, Concept, or Proof), semantic dependencies (Text, Formula, or Both), and different difficulty levels (Low, Medium, and High). Note that the $y$-axes of all plots have the same scale.}
  \label{fig:map}
\end{figure*}

\subsection{Baselines}

For the NTCIR-12 dataset, we compare our scores to the only Transformer retriever reported on this dataset, i.e., MathBERT~\cite{peng2021mathbert}, a BERT model with specialized structure-aware further pretraining.
However, their results should be regarded as an ensemble run because they are generated from reranking a highly effective run produced by Tangent-CFT~\cite{mansouri2019tangentcft}.

For the ARQMath-2 dataset, we select other bi-encoder Transformer runs submitted or reported on the ARQMath-2 main task~\cite{mansouri2021arqmath2review}.
This includes ColBERT runs based on different backbones from the TU\_DBS team~\cite{reusch2021tu_dbs}, using the weights of the original BERT-base, SciBERT~\cite{beltagy2019scibert}, and a Col-ARQBERT pretrained from scratch on the ARQMath corpus.
Furthermore, two additional effective bi-encoder models---CompuBERT~\cite{novotny2021ensembling} from MIRMU and FormulaEmb~\cite{indian2021bert}---are also compared.
The former uses averaged token embeddings of SentenceBERT fine-tuned by minimizing the cosine distance of questions to their accepted or high-ranking answers,
while the latter uses pretrained Transformer embeddings directly for similarity computations.

In our fusion results, we compare top-effective existing systems.
For the NTCIR-12 dataset, we include:
MCAT~\cite{kristianto2016mcat} -- an expensive MIR system that takes on average over 25 seconds per query to run;
the Tangent-S system~\cite{davila2017layout} using low-granularity structure node pairs;
and its successor Tangent-CFT~\cite{mansouri2019tangentcft} based on FastText embeddings of local structures from the SLT representation;
a GNN model for formula retrieval~\cite{song2021GNN};
and finally, MathBERT~\cite{peng2021mathbert}.
However, the two most effective ensemble runs, TanAPP~\cite{mansouri2019tangentcft} and MathAPP~\cite{peng2021mathbert}, are excluded because their linear fusion weights are tuned directly on the complete NTCIR-12 dataset.

For the ARQMath-2 dataset, we include the most effective systems for comparison:\ the MathDowsers primary system~\cite{fraser2018tangentL,ng2020dowsing,ng2021dowsing,ng2021thesis} and the up-to-date Approach0 system.
Additionally, two cross-encoder dense retrievers are included: the TU\_DBS primary retriever based on ALBERT~\cite{reusch2021albert} and  QASim~\cite{mansouri2021dprl}, which combines two Transformers, for question-question and question-answer similarity assessment.
We also consider ensemble systems including the most effective run (WIBC) from the MIRMU team~\cite{novotny2021ensembling} and the official tf--idf and tf--idf+Tangent-S baselines provided in the ARQMath-2 main task.
The tf--idf+Tangent-S baseline is an unweighted average fusion between the results produced by the Terrier system~\cite{Ounis2005Terrier} and a structure-search system, Tangent-S~\cite{davila2017layout}.
In the Terrier pass, \LaTeX{} strings are directly used for retrieval.

\section{Results}

\subsection{Overall Comparisons}
\label{sect:comp}
Evaluation results for Transformer-based dense models are shown in Table~\ref{tab:dense}.
Across both formula-only retrieval (NTCIR-12) and math-aware full-text retrieval (ARQMath-2), our pretrained backbones can generally boost downstream DPR retrieval effectiveness compared to DPR models based on vanilla BERT, row (1), or SciBERT, row (2).
This is presumably because we have further pretrained on more domain-specific data (unlike SciBERT, which also includes scientific text like biomedical articles) with a much larger batch size, i.e., 240 compared to SciBERT's 32 batch size.
According to rows (2)--(4) in Table~\ref{tab:dense}, with pretraining only for 3 epochs, our model reaches a similar level of effectiveness as SciBERT; and more pretraining results in better downstream effectiveness.

Our fully-trained ColBERT model, row (12), achieves the best scores among other Transformer models.
Compared to other ColBERT variants from rows (6)--(8) submitted by the TU\_DBS team, we also achieve higher scores.
Our DPR model, row (11), is generally more effective than previous bi-encoder systems, so it can be considered a cost-effective alternative to ColBERT, since the latter requires a much bigger index.
For ARQMath-2, our DPR model requires a 5.3G index (at full precision), while our ColBERT model requires a 77G index (at half precision).
However, our dense models are not on par with the MathBERT run on the NTCIR-12 dataset, row (5).
This is because MathBERT reranks a highly effective run generated by Tangent-CFT; the latter is directly tuned on complete NTCIR-12 data.

\subsection{Comparisons on Different Topics}

To further investigate the strengths and weaknesses of different architectures for math-aware retrieval, we break down results by different types of topics, e.g., \textit{computation}, \textit{concept}, or \textit{proof}.
Topic categories are labeled by the ARQMath-2 task organizers~\cite{mansouri2021arqmath2review}.

As shown in Figure~\ref{fig:map}, the DPR model, compared to itself, is good at text retrieval but bad at formula retrieval, while ColBERT is the opposite.
The cross encoder (from left to right, the 3rd plot), on the other hand, handles all types of dependencies equally well, and it shows sufficient understanding for easy math question retrieval and proof-related topics.
On the other hand, structure search (the rightmost plot) excels at the calculation category and formula-dependent retrieval, with most categories performing even better than the cross encoder.
This demonstrates that matching formula structure is still crucial for effective math-aware search, especially for formula-heavy content such as the calculation category.

\begin{table}
\small
\centering
\caption{\label{tab:fusion_ntcir}
Comparison of effectiveness on the NTCIR-12 Wiki-Formula dataset. We combine a structure search system (Approach0) with our fully-trained DPR and ColBERT models.
End-to-end fusion weights are tuned via cross-validation.
*/** denotes that the compared row performs weaker than the bottom row (i.e., Approach0 + DPR in end-to-end fusion) at $p<$ 0.05/0.01
level using the two-tailed pairwise $t$-test.
Underlined scores are not involved in any test of significance.
}
\hspace*{-0.1in}
\begin{tabular}{l|ll}
\toprule
    \bf Runs & Part. BPref & Full BPref \\
\midrule
\bf Previous  systems && \\
(1) MCAT \citeyearpar{kristianto2016mcat}
& 0.57 & 0.57 \\
(2) Tangent-S \citeyearpar{davila2017layout}
& 0.59 & 0.64 \\
(3) Tangent-CFT \citeyearpar{mansouri2019tangentcft}
& 0.71 & 0.60 \\
(4) GNN \citeyearpar{song2021GNN}
& 0.54 & 0.63 \\
(5) MathBERT \citeyearpar{peng2021mathbert}
&\bf \underline{0.74} & \underline{0.61} \\[0.75ex]
\bf Structure search baseline && \\
(6) Approach0~\citeyearpar{zhong2019structural,zhong2020accelerating}
& 0.54* & 0.63 \\[0.75ex]
\bf Our reranking && \\
(7) Approach0 + DPR
& 0.48** & 0.53* \\
(8) Approach0 + ColBERT
& 0.52* & 0.56* \\[0.75ex]
\bf Our end-to-end fusion && \\
(9) Approach0 + ColBERT
& 0.63 & 0.65* \\
(10) Approach0 + DPR
& 0.62 &\bf 0.70 \\
\bottomrule
\end{tabular}
\end{table}

\begin{table*}
\small
\centering
\caption{\label{tab:fusion}
Results from the most effective runs of previous systems in ARQMath-2 compared to our method for combining a structure search model (Approach0) with our fully-trained DPR and ColBERT models.
End-to-end fusion weights are tuned via cross-validation.
*/** denotes that the compared row performs weaker than the bottom row (i.e., Approach0 + ColBERT in end-to-end fusion) at $p<$ 0.05/0.01
level using the two-tailed pairwise $t$-test.
}
\hspace*{-0.1in}
\begin{tabular}{l|lllll}
\toprule
    \bf Runs & NDCG' & MAP' & P'@10 & BPref \\
\midrule
\bf Previous systems &&&& \\
(1) MathDowsers Primary \citeyearpar{fraser2018tangentL,ng2020dowsing,ng2021dowsing} &0.434&0.169*&0.211*&0.145** \\
(2) TU\_DBS Primary \citeyearpar{reusch2021tu_dbs} &0.377**&0.158**&0.227*&0.158* \\
(3) DPRL QASim \citeyearpar{mansouri2021dprl} &0.388*&0.146**&0.193*&0.135** \\
(4) MIRMU WIBC \citeyearpar{novotny2021ensembling} &0.332**&0.087**&0.106**&0.069** \\
(5) tf--idf (Terrier) \citeyearpar{mansouri2021arqmath2review} &0.185**&0.046**&0.063**&0.046** \\
(6) tf--idf+Tangent-S \citeyearpar{mansouri2021arqmath2review} &0.201**&0.045**&0.086**&0.048** \\ [0.75ex]
\bf Structure search baseline  &&&& \\
(7) Approach0~\citeyearpar{zhong2019structural,zhong2020accelerating} &0.381**&0.189*&0.234*&0.180* \\[0.75ex]
\bf Our reranking &&&& \\
(8) Approach0 + DPR &0.372**&0.169**&0.235*&0.162** \\
(9) Approach0 + ColBERT &0.383**	&0.182**	&\bf 0.276	&0.181* \\[0.75ex]
\bf Our end-to-end fusion &&&& \\
(10) Approach0 + DPR &0.429*&0.203&0.258&0.189 \\
(11) Approach0 + ColBERT &\bf 0.447&\bf 0.215&0.252&\bf 0.202 \\
\bottomrule
\end{tabular}
\end{table*}

\subsection{Fusion Results}

As shown in Figure~\ref{fig:map}, even a cross encoder (without special structure pretraining) can fall short for formula retrieval compared to the structure search approach.
Nevertheless, we want to learn if dense retrieval can be combined with structure search to further advance structure search effectiveness.

Our fusion results are summarized in Table~\ref{tab:fusion_ntcir} and  Table~\ref{tab:fusion}.
On the NTCIR-12 Wiki-Formula dataset (Table~\ref{tab:fusion_ntcir}),
comparing rows (1)--(5),
our linear fusion runs in rows (9)--(10) outperform others in fully relevant BPref scores.
This shows we can generate a good ranking for highly relevant formulas when linearly combining end-to-end dense retrieval and structure search.
The formula-only reranking in rows (7)--(8) is not beneficial,
but on the other hand, end-to-end fusion in rows (9)--(10) is helpful because dense retrieval can improve recall when structure matching is too strict (more discussion below).

On the ARQMath dataset,
comparing rows (7)--(11) in Table~\ref{tab:fusion} and rows (11) and (12) in Table~\ref{tab:dense}, we see that although the structure search baseline produced by Approach0 alone is generally more effective than dense retrieval models, both DPR and ColBERT can still boost the baseline results.
With the assistance of structure search, we are also able to outperform cross-encoder models shown in row (2) and row (3) in Table~\ref{tab:fusion}.
These cross encoders require costly inference over every candidate pair.
In fact, due to the impractical inference times of cross encoders for the ARQMath dataset, the TU\_DBS team had to limit their candidate pool prior to indexing.
Similarly, the DPRL QASim run adopts a smaller TinyBERT model to practically compute similarities for all candidate pairs in a limited set.

Interestingly, across two datasets, reranking is not helpful in general, other than a precision boost in rows (8)--(9), Table~\ref{tab:fusion}.
This is because the dense rerankers are prone to false positives at formula retrieval compared to structure search, and this is especially the case when a dense retriever is used to rerank a highly effective formula retriever baseline.
We report extra experiments to support this argument in Section~\ref{sect:discussion}.
This indicates that the dense retrievers are only complementary to the structure search approach in a way that helps recall rather than reranking.

\begin{table}
\small
\centering
\caption{\label{tab:fusion_other_than_linear}
Other fusion methods evaluated using the most competitive Approach0 + ColBERT model combination on the ARQMath-2 dataset.
*/** denotes that the compared row performs significantly weaker than the linear fusion at $p<$ 0.05/0.01 level using the two-tailed pairwise $t$-test.
ISR and RRF stand for \textit{Inverse Square Rank} and \textit{Reciprocal Rank Fusion}, respectively.
}
\begin{tabular}{l|llll}
\toprule
\bf Fusion & NDCG' & MAP' & P'@10 & BPref 
\\ \midrule
Borda Count & 0.443 & 0.213 & 0.280 & 0.197 \\
CombSUM & 0.411** & 0.213 & 0.296 & 0.216 \\ 
ISR & 0.433** & 0.203 & 0.263 & 0.191 \\
log-ISR & 0.432** & 0.202 & 0.263 & 0.189 \\ 
RRF ($k=60$) & 0.449 & 0.221 & 0.284 & 0.200 \\
\midrule
Linear & 0.449& 0.217&0.279& 0.204 \\
\bottomrule
\end{tabular}
\end{table}

\section{Discussion}
\label{sect:discussion}
Given that linear fusion is able to produce such good results, a natural question to ask is whether other fusion methods can lead to even better results.
Therefore, we compare popular fusion methods\footnote{We use the \texttt{polyfuse} tool: \url{https://github.com/rmit-ir/polyfuse}} on the ARQMath-2 datasets and our results are summarized in Table~\ref{tab:fusion_other_than_linear}.
In all experiments, we directly choose the best fusion parameters tuned on the ARQMath-2 dataset to obtain an optimistic bound for each method.
Table~\ref{tab:fusion_other_than_linear} shows that linear interpolation is sufficient to generate ``good enough'' results that are not significantly worse (sometimes better) than other popular fusion methods.

We further investigate the reasons why structure search and dense retrieval are highly complementary but not so in the reranking case.
After probing a number of queries where fusion runs achieve much better results, we find that the structure constraint imposed on candidates by Approach0 can fail completely when relevant documents do not share any common substructure in math formulas, especially if the query is formula-centered, while dense retrieval has the capacity to find these relevant documents by matching contextual semantics.
On the other hand, structure search is helpful to dense retrieval in cases where an obviously relevant document is found by matching a candidate formula perfectly.

\begin{figure}
  \centering
  \vspace{-0.2in}
  \includegraphics[width=1\linewidth]{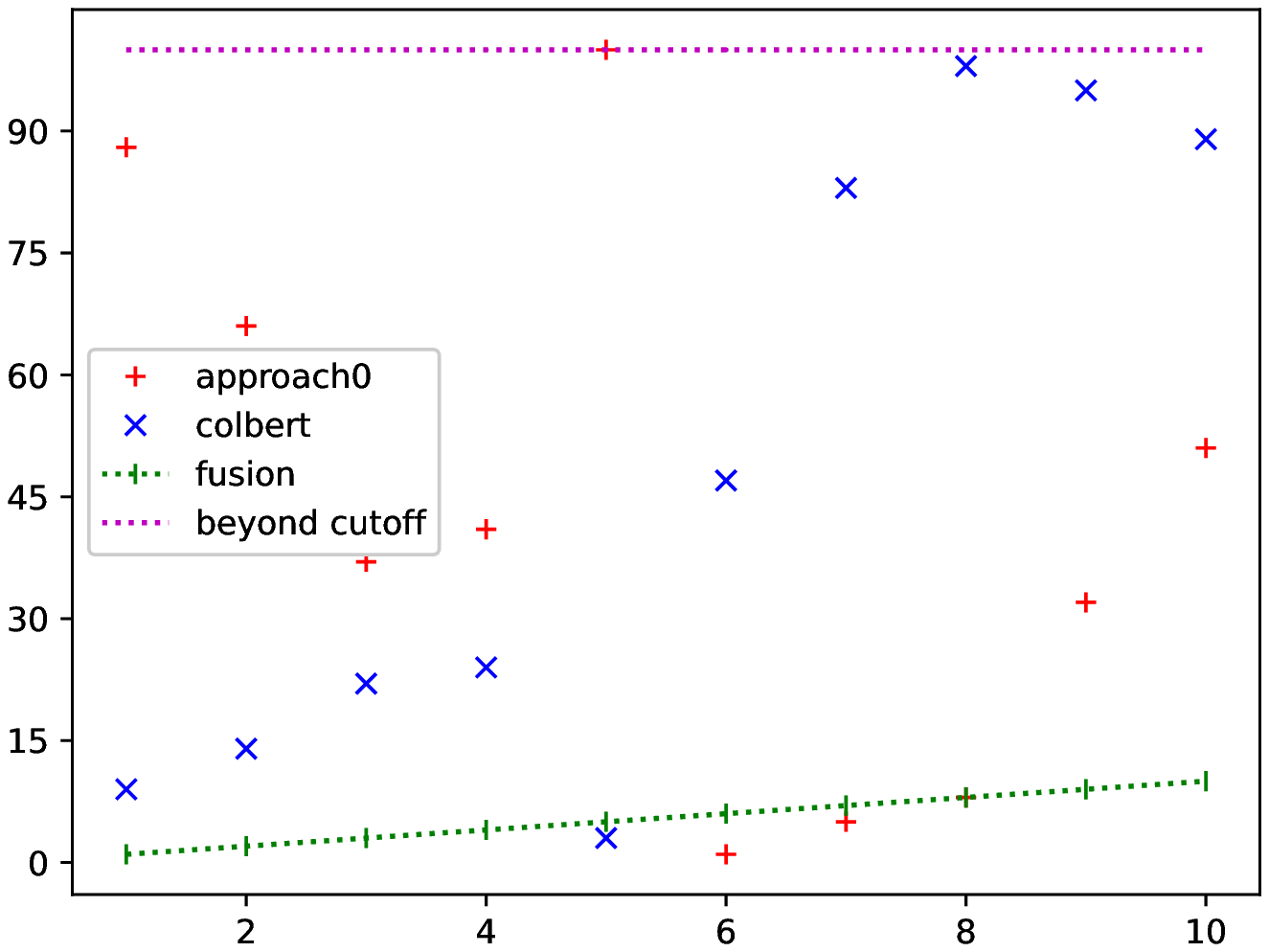}
  \caption{Retrieved document ranks of the Approach0 + ColBERT fusion run (topic A.219, cut off at 10) and their positions in the original runs. The $x$-axis corresponds to the ranks of retrieved documents in the fusion run, the $y$-axis corresponds to the ranks of retrieved documents in the original runs, and each point represents one document.}
  \label{fig:rankchange}
\end{figure}

We illustrate the above statement using the topic where the precision metric has the largest increase after fusion.
Specifically, topic A.287, which gains the most when the fusion run is compared against the Approach0 baseline (P@10 changes from 0 to 0.8), fails for Approach0 because structure match does not occur in relevant documents and the query is formula-centered. 
On the other hand, when compared to the ColBERT run, the fusion run in topic A.219 shows the most gain in precision (P@10 increases from 0.1 to 0.5).
Figure~\ref{fig:rankchange} shows the ranks of top-10 fusion results and their original positions in topic A.219.
By further inspecting in detail, we discover that structure search prevents false positives in dense retrieval.
Specifically, the top-3 dense retrieval hits in topic A.219 contain binomial coefficient notation, e.g., $\sum\limits_{k=0}^r \binom{m}{k}\binom{n}{r-k} = \binom{m+n}{r}$, which looks similar to the query $\binom{n+r+1}{r}=\sum_{k=0}^{r}\binom{n+k}{k}$ but is not equivalent mathematically.
They are ruled out or get lowered in rank in the top-10 final results because their counterparts in structure search are missing (e.g., rank 1st and 2nd results from ColBERT run) or out of sight (e.g., rank 3rd result from ColBERT run), and those ColBERT hits paired with a structure hit in the Approach0 pass stand out.

\section{Conclusions}

Rapid progress in dense retrieval models using deep neural networks has greatly influenced many IR tasks.
In this paper, we provide a thorough evaluation of both token-level and passage-level bi-encoder models in the math information retrieval domain.
Our DPR and ColBERT models adapted to this domain in both pretraining and fine-tuning are made publicly accessible to provide stepping stones for future research.
Our study also highlights the importance of combining structure search with dense retrieval models for better math-aware search.
We show that bi-encoder dense retrieval models alone can be less effective than cross encoders, but when combined with strong structure search methods, they can further improve state-of-the-art effectiveness.
With the huge modeling capacity of dense retrieval models, we believe it is worth exploring other directions for improvements so that we can unleash the potential of deep models in this domain, for example, to better identify similarities in mathematically transformed expressions with different structures.

\section{Limitations}
What our evaluations suggest in this work is to build end-to-end retrievers by combining strict structure search and dense retrieval for a highly effective math-aware search.
We are aware of two limitations:
First, an ensemble of two different end-to-end retrieval systems imposes engineering challenges; the benefit of supporting math-aware search may be offset by the overhead of adding multiple software stacks.
Second, it is unclear how to highlight matching in the case of DPR;
and in the case of ColBERT, it demands larger storage (see Section~\ref{sect:comp}) and requires intensive GPU resources to perform the MaxSim operation over the embeddings of all candidate tokens (see Section~\ref{sect:models}).

\section*{Acknowledgments}
This research was supported in part by the Canada First Research Excellence Fund and the Natural Sciences and Engineering Research Council (NSERC) of Canada.
Computational resources were provided in part by Compute Ontario and Compute Canada.

\bibliography{references}
\bibliographystyle{acl_natbib}

\end{document}